\documentclass[aps,pre,twocolumn,showpacs,superscriptaddress]{revtex4-1}
\usepackage{epsfig}
\usepackage{times}
\usepackage{amsmath}
\usepackage{natbib}
\usepackage{graphicx}
\usepackage{color}
\usepackage[breaklinks=true]{hyperref}
\usepackage{breakcites}
\usepackage{hypernat}
\usepackage{float}
\usepackage{epstopdf}

\begin{document}

\title{Heterogeneity in evolutionary games: an analysis of the risk perception}

\author{Marco A. Amaral} 
\email{marcoantonio.amaral@cpf.ufsb.edu.br}
\affiliation{Universidade Federal do Sul da Bahia - BA, Brazil}

\author{Marco A. Javarone}
\email{marcojavarone@gmail.com}
\affiliation{Department of Mathematics, University College London, London, UK}

\begin{abstract}
In this work, we analyse the relationship between heterogeneity and cooperation. Previous investigations suggest that this relation is nontrivial, as some authors found that heterogeneity sustains cooperation, while others obtained different results.
Among the possible forms of heterogeneity, we focus on the individual perception of risks and rewards related to a generic event, that can show up in a number of social and biological systems.
The modelling approach is based on the framework of Evolutionary Game Theory. To represent this kind of heterogeneity, we implement small and local perturbations on the payoff matrix of simple $2$-strategy games, as the Prisoner's Dilemma. So, while usually the payoff is considered as a global and time-invariant structure, i.e. it is the same for all individuals of a population at any time, in our model its value is continuously affected by small variations, both in time and space (i.e. position on a lattice).
We found that such perturbations can be beneficial or detrimental to cooperation, depending on their setting. Notably, cooperation is strongly supported when perturbations act on the main diagonal of the payoff matrix, whereas when they act on the off-diagonal the resulting effect is more difficult to quantify.  
To conclude, the proposed model shows a rich spectrum of possible equilibria, whose interpretation might offer insights and enrich the description of several systems.
\end{abstract}

\maketitle

\section{Introduction}

The inherent heterogeneity of human beings, as well as that of other animals, reflects in the variety of behaviours observable in groups and communities of individuals. %
However, by using a well defined set of categories to describe human behaviours~\cite{social_psychology}, such as conformity, rationality, zealotry, and so on, the complexity of a social system can be reduced and can be tackled by appropriate models.
The modern field of sociophysics (also called social dynamics)~\cite{galam01, galam03, galam02, loreto01, sen01, perc_bs10, perc_1911, Capraro2018, Sznajd17} represents a class of models whose goal is studying social phenomena and human behaviours in mathematical terms.

Therefore, following an approach similar to that adopted in physics, investigations in sociophysics usually define simple models, based on few degrees of freedom, and solve them by analytical calculations or numerical simulations. This approach is extremely successful, for instance, when studying the motion of particles, and other classic physical systems.

On the other hand, the selection of relevant properties, and the granularity of the corresponding variables, might entail the loss of significant information. 
For instance, when studying human psychology, behaviours, and in general social phenomena, such loss can be so relevant that psychologists and sociologists might be reluctant to accept results obtained by a 'too simple' model, even if based on an elegant set of equations. In saying 'too simple', we refer to those models that oversimplify a real scenario, or phenomenon, becoming unable to represent it properly. It can be worth to mention the case of the voter model~\cite{holley_ap75}, whose limits and potentialities for describing real political scenarios have been reported in~\cite{gracia14}.

Now, let us proceed with our focus on the heterogeneity, also called diversity. Despite being an almost ubiquitous property of social and biological systems~\cite{Sparrow1999}, for the sake of simplicity many models can safely neglect it. On the other hand, sometimes further efforts to represent heterogeneity led to findings more valuable than expected (see for instance~\cite{santos_jtb12}). Also, since this property can refer to one or more aspects of a system, seeing it as beneficial or detrimental becomes a fairly relative matter.
For instance, networks~\cite{barabasi01} can be heterogeneous or homogeneous in relation to their topology, like scale-free~\cite{barabasi02} and classical random graphs~\cite{erdos_pmd59}, respectively. Here, both the transmission of information and the viral spreading are slower in networks with homogeneous topology. So, in the first case, heterogeneity is clearly beneficial, while in the other is detrimental.
Interesting observations arise also considering animal cell replication, which can be realized by mitosis and meiosis. Notably, the latter increases the genetic diversity (i.e. heterogeneity)~\cite{genetics01} of a population.
In business and science, heterogeneity of skills can make a group more effective (e.g.~\cite{Armano2017}), in ecological systems it could affect the size and the stability of a group (see~\cite{Javarone2017}), and in social systems heterogeneity (e.g. cultural differences) might constitute the cause of a conflict~\cite{social01}.
Recently, the existence of a deep relationship between cultural differences and cooperation in human societies has been experimentally confirmed~\cite{handley20}.
It is worth to mention that cooperation still represents an important scientific challenge, and despite a huge literature focused on its dynamics (e.g.~\cite{rand_tcs13, Perc2017,buchan_pnas09, Gomez-Gardenes2007,duh01}), a lot of questions around it still require a proper answer. Hence, clarifying further whether and how some forms of heterogeneity can support cooperative behaviours might be relevant.

To this end, we concentrate on the heterogeneity of the individual perception of a risk (or reward) associated with a generic event, i.e. we study an aspect of humans and, probably also of other animals, that can manifest in a variety of common scenarios. In addition, since our investigation aims to understand the relationship between this psychological aspect and cooperative behaviours, we use the framework of Evolutionary Game Theory (hereinafter EGT)~\cite{Nowak2006, Hofbauer1998}. The latter allows studying social dynamics, biological phenomena, and many other evolutionary systems~\cite{Szabo2007}. In addition, heterogeneity has already been analysed under its lens and a variety of results, sometimes even controversial, has been obtained~\cite{gracia-lazaro_pnas12, perc_pre08, tanimoto_pre07b, Zhang2013}. In our view, an increasing interest in the relationship between heterogeneity and cooperation reflects its potential relevance in many contexts and domains. This can also be observed in the influx of recent works trying to find a general framework to explain the effects of perception diversity~\cite{Szolnoki2019, Su2019, Stollmeier2018, Szolnoki2018, Alam2018, Hilbe2018, Qin2017a, Javarone2016b, Tanimoto2016, Amaral2016, Yakushkina2015, wang_z_pre14b, Zhang2013, santos_jtb12, anh_tpb12}.

Thus, motivated by a challenging debate, we analyse the heterogeneity resulting from varying the profits that rational individuals try to achieve in simple games. In saying rational~\cite{Javarone_book}, we assume that all individuals act to maximise their profit, so that any variation of the payoff structure can actually influence their actions.
Accordingly, in the proposed model, individuals play games like the Prisoner's Dilemma, whose payoff matrix is affected by small random perturbations. 
As one can note, here the payoff is not a global and constant property of the system, since small random perturbations can last a single interaction, and can affect a single individual at a time. In doing so, we represent a population whose perception of risk/reward is not defined by a static and global parameter but, instead, by a value that fluctuates around an average.

We remark that many real scenarios involve people having a different risk perception of the same event. Notably, from financial markets to scientific organisations, the psychology of risk perception plays a fundamental role, with a spectrum of behaviours (e.g. group polarisation) that can become even dangerous. A relevant case can be found in~\cite{whyte89}, which describes the failure of the NASA Challenger project as an example of a dangerously reduced perception of risk.
Hence, beyond being a common aspect of the real world, heterogeneity of risk perception is also quite relevant and difficult to predict or quantify.
We emphasise that the perturbation approach here developed finds connections also with new researches on potential games~\cite{Szabo2016a}, that use the operator formalism from quantum mechanics to relate the payoff matrix with the Hamiltonian of a system~\cite{Nakamura2019}.

The next sections describe in detail the proposed model, show the results of numerical simulations and, eventually, discuss our findings and future developments.
\section{Model}
\label{sec:Model}

In the proposed model, we consider two-strategy games whose agents can either cooperate (C) or defect (D). Mutual cooperation yields a payoff $R$ (reward), while mutual defection yields $P$ (punishment). Also, a defector receives a payoff $T$ (temptation) when interacts with a cooperator that, in turn, receives a payoff $S$, known as the Sucker's payoff~\cite{Szabo2007}.
The payoff matrix combines in a single structure all the above-mentioned payoffs so, depending on their value, it relates to a specific game (e.g. the Prisoner's Dilemma). Usually, this matrix is constant over time and it contains the same entries for all agents. In order to represent the heterogeneity of risk perception, we introduce small perturbations on this matrix.

Notably, such perturbations act at the local level hence, at a same instant, two individuals might have a slightly different payoff matrix, as well as one individual can find two slightly different payoff matrices in two different game iterations.
In particular, perturbations (i.e. $\epsilon$) are implemented via a uniform distribution with range $D$ (our control parameter) and they are not cumulative. 
Moreover, it is important to add that the resulting heterogeneity has zero average, i.e. perturbations should actually give a null effect in the long term.

\begin{figure}
  \includegraphics[width=5.7cm]{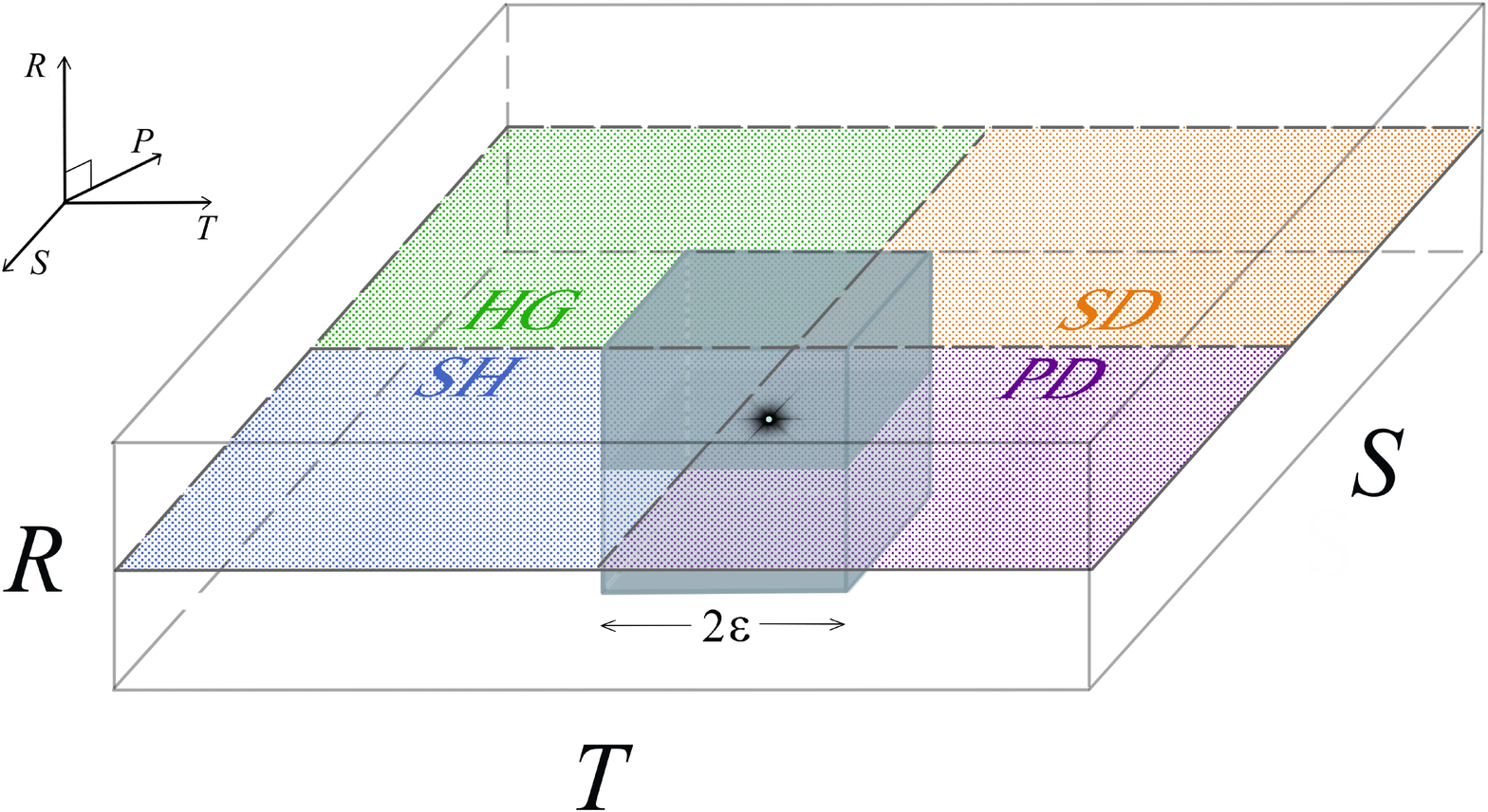}
  \caption{$T\times S \times R \times P$ (4 dimensional) parameter space with $R=1,P=0$, spanning four classes of games. The $P$ dimension is represented only in the Cartesian coordinate system (outside the parameter space, on the left-hand side). The payoff fluctuation can act over $[T,S,P,R]$ simultaneously and uncorrelated. This makes any given game able to fluctuate around a (4D) box with $2\epsilon$ size edges, centered on the original game. Note that local fluctuations can lead agents to (locally) play different classes of games depending on the fluctuation strength.}
  \label{diagram}
\end{figure}

We consider three different configurations: FP (Full Perturbation, i.e. acting on all payoff entries), MDP (Main Diagonal Perturbation, i.e. acting only on the main diagonal of the payoff matrix) and ODP (Off-Diagonal Perturbation). Note that, by doing so, there are no asymmetries between the strategies regarding the perturbations, i.e. for any configuration both strategies, C and D, will always have at least one of their possible payoff values perturbed. The considered configurations do not cover the whole spectrum of possibilities as, for instance, another configuration could be based on perturbations acting only in the $R$ and $S$ entries, and it would result in a population where only cooperators would be affected.
The first configuration (FP) has the following payoff matrix:
\begin{equation}
\begin{array}{c c} &
\begin{array}{c c} C~~ & ~~D \\
\end{array}
\\
\begin{array}{c c}
C \\
D
\end{array}
&
\left[
\begin{array}{c c }
R+\epsilon _R & S+\epsilon _S \\
T+\epsilon _T & P+\epsilon _P
\end{array}
\right]
\end{array}
\end{equation}

the MDP configuration is defined as
\begin{equation}
\begin{array}{c c} &
\begin{array}{c c} C~~ & ~~D \\
\end{array}
\\
\begin{array}{c c}
C \\
D
\end{array}
&
\left[
\begin{array}{c c }
R+\epsilon _R & S \\
T & P+\epsilon _P
\end{array}
\right]
\end{array}
\end{equation}

eventually, the ODP configuration reads
\begin{equation}
\begin{array}{c c} &
\begin{array}{c c} C~~ & ~~D \\
\end{array}
\\
\begin{array}{c c}
C \\
D
\end{array}
&
\left[
\begin{array}{c c }
R & S+\epsilon _S \\
T+\epsilon _T & P
\end{array}
\right]
\end{array}
\end{equation}

\noindent where $T\in[0,2]$ and $S\in[-1,1]$. Without loss of generality, we set $R=1$ and $P=0$ \cite{Perc2006a}. Although $[R,P]$ are fixed, the perturbation can make said values fluctuate for each individual and at each time (i.e. iteration). 
The parametrization adopted in the above payoff matrices spans four different classes of games in the $[T,S]$ parameter space: prisoner's dilemma (PD), snow-drift (SD), stag-hunt (SH), and harmony games (HG)~\cite{Szabo2007, perc_bs10} ---see Figure~\ref{diagram}. We stress that the actual parameter space has 4 dimensions, and the perturbations will act on all four parameters $[T,S,P,R]$ in an uncorrelated manner. 

It is a well known result that the Nash Equilibrium of a game is not changed by the sum of a constant term in the payoff matrix nor by its multiplication by a constant~\cite{Szabo2007, Hofbauer1998}. However, it is important to remember that we are adding random numbers in each payoff entry, and not global constants. So, the actual individual payoff value will change in time and for each agent.
Accordingly, although the average value of the perturbation is zero, locally the game class can change from time to time~\cite{Perc2006, tanimoto_pre07b, hofbauer_jet07}. Even so, the average game will stay centered in the chosen initial $[T,S]$ parameters.
For the population dynamics, we implement the usual imitative update rule weighted by the Fermi distribution~\cite{Javarone_book}, in a spatially distributed population with the square lattice topology. At each time step, one agent, $i$, updates its strategy by comparing its payoff with one randomly chosen first neighbour, $j$. Then, agent $i$ adopts the strategy of agent $j$ with probability
\begin{equation}\label{imitateeq}
p(\Delta u_{ij})=\frac{1}{ 1+e^{-(u_{j}-u_{i})/k} },
\end{equation}
\noindent where $k$ is the irrationality level \cite{Szabo2007}, and $u_{i}$ represents the payoff of agent $i$. We set $k=0.1$ for all simulations unless stated otherwise.
\section{Results}
\label{sec:Results}
We begin with a simplified analytical model of a mixed game setting and try to generalise it to the perturbation payoff scenario. As shown in~\cite{Amaral2015}, the random mixing of two fixed different games (i.e. payoff matrices) in the context of the well-mixed population should be identical to the evolution of a population using the average payoff matrix in the mean-field approximation (note that this is not necessarily true for spatially distributed systems). 
We can generalise this result for the case of infinite discrete games randomly mixed, which represents our model of perturbed payoffs. Suppose that at a time $t$, a system is characterised by the strategy configuration $\{s\}=\{s_1,s_2,...s_N\}$, where $s_i$ is the strategy of agent $i$, and by the payoff matrix assignment configuration $\{g(t)\}=\{g_{12}(t),g_{13}(t),\ldots\}$. Note that the variable $g_{ij}(t)$ indicates which game agents $i$ and $j$ are playing at time $t$. 
In our specific case, $g_{ij}(t)$=$G+\epsilon_{ij}(t)$ where $G$ is the fixed usual payoff matrix, and $\epsilon_{ij}(t)$ is the perturbation that is randomly drawn and it has zero average.
Let $P(\{s\},\{g\},t)$ be the probability to find the system in the configuration $\{s\}$ and $\{g\}$ at time $t$. The time evolution of this system is given by the master equation:
\begin{eqnarray}\nonumber
\label{mastereq}
\frac{d}{dt}P(\{s\},\{g\},t)=\sum_{\{s\}',\{g\}'} P(s',g',t)W(s',g'\rightarrow s,g) \\ -P(s,g,t)W(s,g\rightarrow s',g')~~,
\end{eqnarray}
\noindent where $W(s',g'\rightarrow s,g)$ is the transition rate from the state $\{s\},\{g\}'$ to $\{s\},\{g\}$.

Supposing that the strategy variables $\{s\}$ are statistically independent of the game variables $\{g\}$, and that the transition rates are linear functions of the payoffs, we can use the main result from~\cite{Amaral2015} to obtain:
\begin{eqnarray}
\left\langle \frac{d}{dt}P(\{s\},\{g\},t) \right\rangle _{g} =\frac{d}{dt}P(\{s\},\left \langle g \right \rangle ,t).
\end{eqnarray}
That is, the average evolution of a population playing a set $\{g\}$ of randomly mixed games will undergo the usual evolution of a single population playing the equivalent average payoff matrix $\langle g \rangle$. 

We stress that this initial result only applies to the mean-field approximation. Nevertheless, it is useful for comparison with our model, since the average payoff matrix of the perturbed game is equal to the unperturbed matrix. Analytical results predict that, on average, small random perturbations should not change the evolution as long as the population is interacting without spatial structure and the update rule is linear. This is also in line with recent results reported in~\cite{Zhou2018}. Our simulations using a well-mixed topology presented some small changes in the perturbed population evolution, with the perturbations enhancing cooperation. In contrast, the outcomes for the square lattice show strong and diverse emergent effects caused by the perturbations. We proceed by presenting the main results for the square-lattice topology.

To implement the numerical simulations we adopt an asynchronous Monte-Carlo protocol with a square lattice, Von Neumann neighbourhood and periodic boundary conditions with $N=10^{4}$ agents. We wait for the system to reach a dynamic equilibrium (around $10^4$ MCS's) and average the values over the final 1000 steps. This is repeated for $50$ different samples with the same parameters.

\begin{figure}
  \includegraphics[width=7cm]{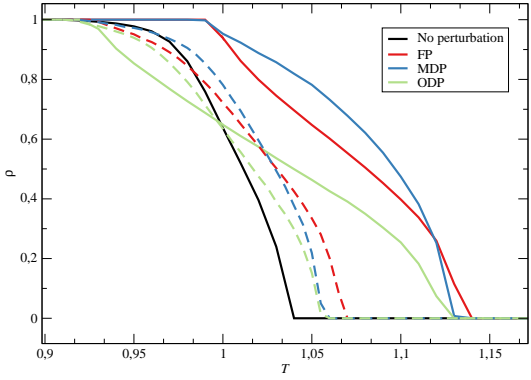}
  \caption{Average final cooperation level as a function of $T$ for the three different configurations, Full Perturbation (FP), Main Diagonal Perturbation (MDP) and Off-Diagonal Perturbation (ODP). Dashed lines represent small perturbation ($D=0.1$) and continuous lines represent strong perturbation ($D=0.3$). Even small perturbations can sustain cooperation past the extinction point and larger noise strength can greatly enhance cooperative behaviour. Note however that the effects of each configuration are slightly different.}
  \label{Tvar1}
\end{figure}

We begin by focusing on the final fraction of cooperators ($\rho$) near the phase transition in $T_c=1.04$. For the sake of simplicity, we present the outcomes related to the weak prisoner's dilemma case (i.e. $S=0$). Figure~\ref{Tvar1} shows results achieved by small (i.e. $D=0.1$) and medium ($D=0.3$) perturbations, compared to the unperturbed case ($D=0$). The main differences between the perturbed configurations and the unperturbed one happen in the region around $0.9<T<1.15$.
From Figure~\ref{Tvar1}, we see that all three kinds of perturbation settings can enhance the range of $T$ where cooperation survives. Even more, we see that for strong perturbations the MDP model seems to be the most efficient in doing so, being followed by the FP model. The general effects of the three different perturbation configurations become more and more similar as $D\rightarrow0$. This effect was consistently observed for other $D$ values. It is also worth pointing out that the extinction point of cooperation is very similar for all three configurations.

\begin{figure}
  \includegraphics[width=7cm]{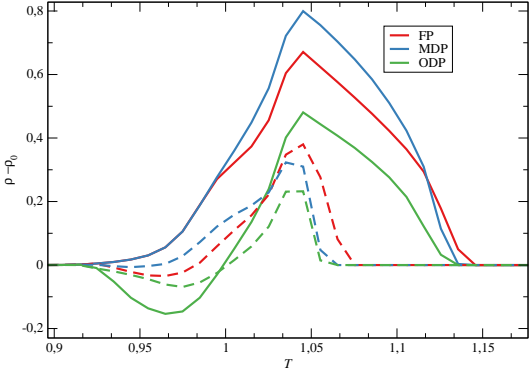}
  \caption{Difference between the final cooperation level of the perturbed ($\rho$) and non-perturbed ($\rho_0$) cases as a function of $T$. By analisying $(\rho-\rho_0)$ we can filter off the effect of increasing $T$, leaving just the perturbation effect. Dashed lines represent small perturbation ($D=0.1$) and continuous lines represent an average perturbation ($D=0.3$). While the FP and MDP settings increase cooperation, the ODP has a non-trivial effect, decreasing cooperation if $T<1$ (Harmony-Game region).}
  \label{Tvar2}
\end{figure}

Figure~\ref{Tvar2} presents the difference between the final cooperation level of the perturbed ($\rho$) and the unperturbed ($\rho_0$) case. Again, for simplicity, we present only the cases for $D=0.3$ and $D=0.1$. We report that simulations with intermediate $D$ values also present curves that change continuously. We can observe that the overall effect of the perturbation is to increase cooperation. Nevertheless, the ODP model can hinder cooperation for the small region $0.9<T<1$. In Figure~\ref{Tvar2}, it is clear that the MDP model can have a greater impact on cooperation. What is more, if we consider that the FP model is the sum of the MDP and ODP perturbations, it becomes clear that the small negative values of $\rho-\rho_0$ observed for the weak perturbations in the FP model (red dashed line) in the region $0.9<T<1$ are due to the off-diagonal perturbations. We also note that while the FP model is the direct sum of the MDP and ODP model in terms of perturbations, the final cooperation level in the FP model is not just the sum between the final cooperation level of the ODP and MDP models. That is, as expected, the final cooperation behaves non-linearly in the types of perturbations implemented.

\begin{figure}
  \includegraphics[width=7cm]{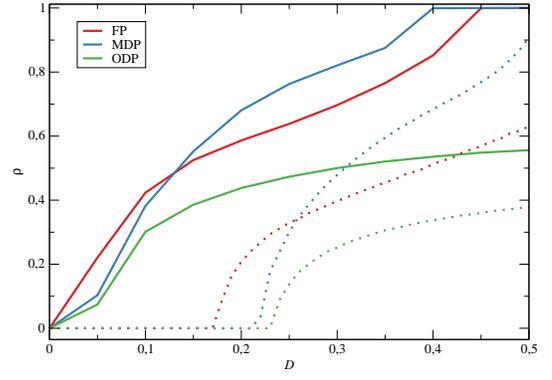}
  \caption{Average cooperation level as a function of $D$ for the three configurations using $T=1.04$ (continuous lines) and $T=1.1$ (dotted lines). The cooperation enhancing effect increases monotonously with the perturbation strength. Such effect mostly pronounced in the cooperation extinction point $T_c=1.04$.}
  \label{deltavar}
\end{figure}

In figure~\ref{deltavar} we present the final cooperation level as we vary the perturbation strength for $T=1.04$ (continuous lines) and $T=1.1$ (dotted lines) in the three configurations. This allows us to continuously visualise the effect of the perturbation strength. Note that for both $T$ values, the FP model generates more cooperation for small perturbation effects. However, the MDP model quickly overcomes this difference and it becomes the more efficient type of perturbation regarding the promotion of cooperation. The specific turning point varies with $T$ in a nontrivial manner. Also, note that while the ODP configuration has a very shallow slope, the other two configurations keep increasing $\rho$ at a fast rate as $D$ increases. We see again that the perturbation in the off-diagonal elements of the payoff matrix is of little effect when compared to the main diagonal terms.

\begin{figure}
  \includegraphics[width=7cm]{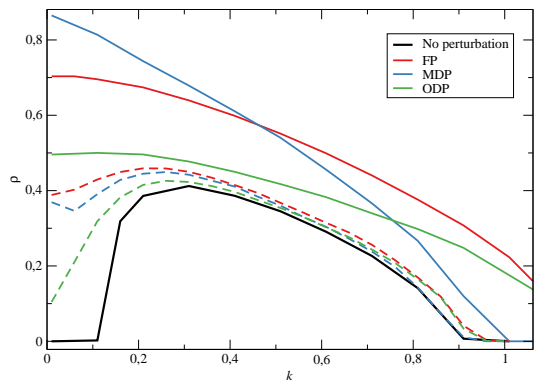}  
  \caption{Final cooperation level versus the irrationality level $k$ for the three different configurations at $D=0.3$ (continuous lines) and $D=0.1$ (dashed lines). The solid black curve shows the unperturbed model, $D=0$. Notice that while fluctuations originating from varying irrationality ($k$) can boost cooperation, they are not correlated to the fluctuations resulting from payoff diversity ($D$).}
  \label{vark}
\end{figure}

Despite our main focus on the behaviour of rational individuals, for the sake of completeness, here we briefly present the results achieved by continuously varying the level of irrationality (controlled by the associated temperature $k$). The results can be seen in Figure~\ref{vark}. Notably, increasing $k$ (i.e. adding some irrationality) one observes random fluctuations in the imitation process, however it is important to realise how differently that variation affects our population in comparison to the effects resulting from random perturbations in the payoff.
The payoff perturbation can increase cooperation for all the studied values of $k$, although, after a peak value, this effect begins to diminish as irrationality keeps increasing. We can also see that this increase is proportional to the perturbation strength. On the other hand, it is more difficult to compare the effects of each perturbation configuration. We see that for low $k$ values, the most influential is the MDP configuration, followed by the FP and lastly the ODP. Nevertheless, for high $k$, their effect is inverted.
In the context of the evolution of cooperation, scenarios with low irrationality are more interesting, since they lead to more cooperation and as $k$ becomes much larger than $1$ the system just behaves randomly. In the region of $k \in [0,1]$, we see again that the diagonal perturbations are the most efficient route to promote cooperation. For low perturbations, however, the FP is a little more efficient.

We stress that the random perturbation effects on payoff entries lead to different outcomes than randomness arising from the irrationality level $k$. Specifically, we ran simulations for a wide range of $k$ (i.e. $k\in(0,3]$) and while the perturbation effects remained qualitatively similar, the pure effects of varying $k$ are very different from those of the payoff perturbation. This can be more clearly seen by comparing Figures~\ref{vark} and~\ref{deltavar}. Further interesting analyses, on the effects of irrationality, have been reported also in~\cite{Szabo05, Vukoc01}.

It is worth mentioning that payoff perturbation changes population dynamics for very short Monte-Carlo times. Remarkably, even earlier than 100 Monte-Carlo steps the perturbation can change the population evolution. This effect has been observed in all three configurations. We also observe that lattice snapshots were not useful in characterising the population behaviour with regard to payoff perturbations. That is, while perturbations increase the overall cooperation level, snapshots of the perturbed and unperturbed populations are largely similar, with the only difference being in a greater density of cooperators for all perturbed configurations.
\section{Conclusions}
\label{sec:Conclusions}
In social and biological systems, and in many other contexts, heterogeneity often shows up in different forms. Understanding the way it contributes to a specific process is not always intuitive or immediate. Also, the inherent heterogeneity of a system is frequently neglected for practical modelling reasons, and often this choice turns to be convenient. However, the relationship between heterogeneity and cooperation is still a matter of open debate.

Notably, previous investigations addressing other forms of heterogeneity (e.g.~\cite{schreiber_tpb13,santos_n08, huang15, santos_pnas06, Amaral2018}) suggested that its relationship with cooperation is absolutely relevant, whereas other authors obtained results leading to an opposite, or different, views~\cite{perc_njp11,gracia-lazaro_pnas12}. Therefore, this variety of findings turned the relationship between heterogeneity and cooperation into a stimulating debate. At this point, we deem relevant to mention also a very recent investigation~\cite{handley20}, related to large-scale forms of cooperation, reporting a clear connection between human cooperation and cultural differences.
Looking at this scenario, in our view it is definitely relevant to identify the system parameters (and conditions) whose heterogeneity supports the emergence of cooperative behaviours, and trying to compare models that lead to (maybe only apparent) conflictual outcomes. Notably, even a small variation of a configuration can change drastically an equilibrium.

So a further, maybe somehow speculative, generalisation of our view can read as the need to identify parameters whose small variations can affect the strategy equilibrium, in the same way as in Physics one aims to find quantities whose variation does not change the equations of motion. 
For the sake of clarity, we specify that while the Noether theorem is focused on the relation between symmetry and conservation, in this context, the opposite seems to be definitely more interesting, i.e. the need to identify those variables whose heterogeneity leads to a different equilibrium.

Let us now go back to the main question, i.e. is heterogeneity a beneficial ingredient for stimulating cooperative behaviours?
Notably, we considered the heterogeneity of reward and risk perception for two main reasons. First, risk perception is an individual aspect of humans, and probably also of other animals, that influences the way individuals act in the real world. Second, the results can find applications also in other domains, as biology and clinical research~\cite{Stollmeier2018}.
Thus, before discussing the main achievements, we deem important to begin with a sort of claim, that is, our findings corroborate the idea that heterogeneity might influence cooperation. Then, how such influence takes place is something to clarify properly.
Let us recall that for representing this property (i.e. heterogeneity), we added small and local perturbations on the payoff matrix of $2$-strategy games. Then, we observed a nontrivial, and partially unexpected, spectrum of outcomes. 
In particular, we analysed three cases, i.e. perturbations acting on the whole payoff matrix, acting on the main diagonal, and on the off-diagonal, separately.
The effect of these perturbations has to be evaluated in the proximity of the phase transition of the system, i.e. within a narrow set of values of the payoff matrix since far from this critical area the evolution of strategies is dominated by average payoff values and is poorly affected by small variations.
It is also worth to remark that perturbations act independently on each single entry of the payoff matrix, i.e. perturbations are generated by four independent uniform distributions.
So, in the first case (i.e. perturbations acting on the whole matrix), cooperation is clearly supported by heterogeneity. Then, considering each diagonal individually, i.e. the main diagonal and the off-diagonal, we found some nontrivial outcomes. In particular, perturbations acting only on the main diagonal are more effective in fostering cooperation than those acting on the whole matrix.
On the other hand, perturbations acting only on the off-diagonal support defection at the beginning of the phase transition, while support cooperation after a given value.
Therefore, the former case shows an unexpected behaviour, and we deem it also deserves further investigations to be fully clarified.
According to the results of the proposed model, the role of heterogeneity in the dynamics of evolutionary games is definitely nontrivial and difficult predict. Also, as in other systems (e.g.~\cite{Javarone2013a}), individual perception can affect the dynamics of a population.

To conclude, we provide our personal intuitions on the achieved results. Making some abuse in the terminology, we observe that small perturbations on the main diagonal act on 'homogeneous groups' (i.e. $C-C$ and $D-D$), while those on the off-diagonal act on 'heterogeneous groups' (i.e. $C-D$ and $D-C$). 
So, as confirmed also in previous studies (e.g.~\cite{Hilbe2018,Amaral2016}), groups of cooperators are able to obtain positive feedback loops of system fluctuations. The same benefits cannot be obtained by groups of defectors as, by definition, the latter never contribute to their community.
This observation can be related to the results reported in~\cite{Javarone2016b} and in~\cite{meloni09}, where it has been shown that a population of random walkers has a higher probability to cooperate if agents 'move slowly' in a continuous space. In particular, as in the proposed model small perturbations have been more effective in the group $C-C$, temporary groups of cooperators (i.e. whose topology entails $C-C$ interactions) support each other and do that better than groups of defectors, that cannot enjoy a free lunch.

Finally, analysing the dynamics of evolutionary games through the lens of perturbations, in the payoff matrix, could become useful both for modelling other kinds of scenarios, and for making connections with the recent works on potential games, and their description based on Hamiltonian functions~\cite{Nakamura2019,Szabo2016a}. For instance, payoff perturbations could be embedded into an Hamiltonian by following an approach similar to that used in perturbation theory, e.g. splitting the Hamiltonian function in two components, one unperturbed and the other containing the additional (perturbative) term.

\begin{acknowledgments}
This research was supported by the Brazilian Research Agency CNPq (proc. 428653/2018-9).
\end{acknowledgments}


\end{document}